\documentclass[12pt]{iopart}
\usepackage{graphicx}
\usepackage{iopams}
\usepackage{color}

\begin{document}
\title{Stochastic gene expression conditioned on large deviations}

\author{Jordan M Horowitz$^1$ and Rahul V Kulkarni$^2$}

\address{$^1$ Physics of Living Systems Group, Department of Physics, Massachusetts Institute of Technology, 400 Technology Square, Cambridge, MA 02139}
\address{$^2$ Department of Physics, University of Massachusetts Boston, Boston, MA 02125}
\begin{abstract}
The intrinsic stochasticity of gene expression can give rise to large
fluctuations and rare events that drive phenotypic variation in a
population of genetically identical cells. Characterizing the
fluctuations that give rise to such rare events motivates the analysis of
large deviations in stochastic models of gene expression. Recent
developments in non-equilibrium statistical mechanics have led to a
framework for analyzing Markovian processes conditioned on rare events
and for representing such processes by conditioning-free {\em driven}
Markovian processes. We use this framework, in combination with
approaches based on queueing theory, to analyze a general class of
stochastic models of gene expression. Modeling gene expression as a
Batch Markovian Arrival Process (BMAP), we derive exact analytical
results quantifying large deviations of time-integrated random
variables such as promoter activity fluctuations.  We find that the
conditioning-free driven process can also be represented by a BMAP
that has the same form as the original process, but with renormalized
parameters. The results obtained can be used to quantify the
likelihood of large deviations, to characterize system fluctuations
conditional on rare events and to identify combinations of model
parameters that can give rise to dynamical phase transitions in system
dynamics.
\end{abstract}

%

\maketitle

In stochastic systems, it is often of great interest to characterize the fluctuations that give rise to rare events. 
For example, a recurring theme in current biological research is rare events leading to phenotypic variation in clonal cells~\cite{raj2008nature}. Prominent examples
include latency in HIV-1 viral infections~\cite{weinberger2005stochastic}, sporulation in bacteria~\cite{mirouze2011fluctuations}, and reversible drug tolerance~\cite{sharma2010chromatin} in subpopulations of cancer cells. In
several cases, the corresponding rare phenotypic transition is primarily driven by the intrinsic stochasticity of gene
expression. 
These observations provide strong motivation for analyzing rare large deviations in stochastic models of gene expression \cite{assaf2011determining,newby2012isolating}.

The development of a framework for analyzing rare events in stochastic gene expression needs to take into account 
multiple factors. Single-cell experiments indicate complex mechanisms underlying 
bursting ({\it i.e.}\ long periods of inactivity punctuated by shorter periods 
of activity) in gene expression~\cite{suter2011mammalian}, motivating
the study of general stochastic models with
multiple promoter states~\cite{sanchez2011effect,zhang2014promoter}. 
Furthermore, the random variables whose rare fluctuations are of interest can be diverse, ranging from
promoter activity fluctuations~\cite{mirouze2011fluctuations} to the fraction of time spent in
specific promoter states~\cite{weinberger2005stochastic}. 
These observations motivate the analysis of a general class of gene expression models that can accommodate
arbitrary complexity in promoter dynamics and bursting. The
development of an analytical framework for rare events in such models
can be used to address several questions of current interest: (1) How
do combinations of underlying model parameters control the likelihood
of rare events? (2) How can we characterize fluctuations in the system
{\em conditioned} on the occurrence of a rare event?  (3) Can we
determine the changes in dynamical model parameters that mimic the
effects of rare fluctuations?

Recent developments in nonequilibrium statistical mechanics using
large deviation theory provide a framework for addressing such
issues~\cite{Lecomte2007,Touchette2009,Touchette2013}. In particular, considerable interest has
focused on deriving conditioning-free Markov processes, termed
{\em driven} processes, whose statistics reproduce the fluctuations of the original Markov
process conditioned on the occurrence of a rare event~\cite{Jack2010,Chetrite2013}. 
So far, nontrivial examples which explicitly characterize the driven process for infinite dimensional systems 
have been limited. Furthermore, it is of interest to determine systems for which the stochastic generator for the driven process has the same structure as the original process \cite{torkaman2015effective}.
In the following, we combine large deviation theory framework with tools from queueing theory~\cite{williams2016stochastic,arazi2004bridging}, to obtain analytical formulas for the statistics of rare events in a general class of stochastic models of gene expression.
We find that the conditioning-free Markov process that generates rare fluctuations generically maintains the same structure as the original Markov process.

\emph{Model.---}
\begin{figure}[t!]
\centering
\includegraphics[scale=.75]{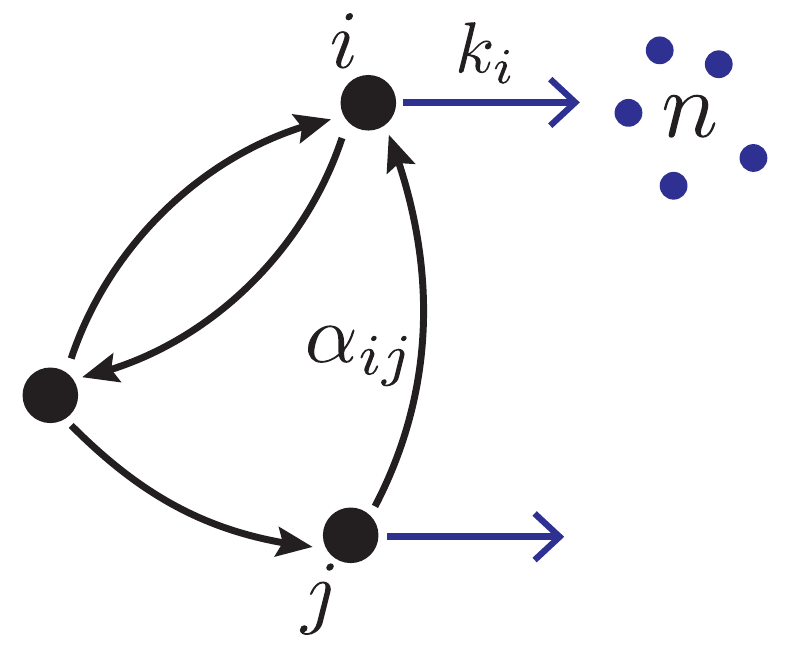}
\caption{Schematic of generic promoter model for bursty mRNA transcription.  The promoter makes random transitions among its $N=3$ states with rates $\alpha_{ij}$.  Bursts of mRNA of size $n$ are drawn from a state-dependent burst size distribution $b_{i}(n)$ and arrive with rate $k_i$.}
\label{fig:promoter}
\end{figure}
We consider gene expression from a promoter
with $N$ internal states, labeled $i=1,\dots,N$, as illustrated in \fref{fig:promoter}.  
The promoter makes random transitions among
its $N$ states switching from $j\to i$ with rate
$\alpha_{ij}$.  
In each state, bursts of gene expression leading to
the production of mRNAs occur with rates $k_i$, and burst sizes $n$
drawn from a state-dependent distribution $b_i(n)$. 
In the limit that protein degradation rates are much smaller than mRNA
degradation rates, a widely-used approximation involves assuming that
proteins are created in random instantaneous bursts from each
mRNA~\cite{friedman2006linking,shahrezaei2008analytical,elgart2011connecting,jia2011intrinsic}. 
Given the validity of this `bursty protein synthesis'
approximation, the model considered in \fref{fig:promoter} can
also be used to represent gene expression at the level of proteins.

The statistical state of the system is specified by a vector ${\bf p}_t(M)=\{p_t^i(M)\}_{i=1}^N$ whose elements are the probabilities $p_t^i(M)$ for the promoter to be in state $i$ at time $t$ having produced a total of $M$ mRNAs. 
The dynamics is an example of what is known in queueing theory as a Batch Markovian Arrival Process (BMAP)~\cite{cordeiro2011batch}, whose evolution is specified by the master equation
\begin{equation}\label{eq:master}
\dot{\bf p}_t(M) = \sum_{M^\prime} {\bf D}_{M-M^\prime}{\bf p}_t(M^\prime),
\end{equation}
where the $N\times N$ matrix ${\bf D}_0$ is the transition rate matrix for intra-promoter transitions, whereas ${\bf D}_n$ ($n \ge 1$) is an $N\times N$ diagonal matrix whose nonzero elements specify the rate of creating $n$ mRNA in a burst:
\begin{equation}\label{eq:D}
D_0^{ij}= 
\left\{\begin{array}{cc}-\sum_j\alpha_{ij}-k_j, & i=j \\
\alpha_{ij}, & {i\neq j}\end{array}\right.;\quad D_n^{ij}=k_j b_j(n)\delta_{ij}.
\end{equation}
The diagonal elements $D^{ii}_0$ enforce probability conservation.
While the representation of the dynamics in terms of the matrices ${\bf D}_n$ is convenient, the full dynamics occur on the infinite-dimensional space spanned by the states $(i,M)$.
Thus, the generator matrix for the master equation \eref{eq:master} is infinite dimensional, 
\begin{equation}\label{eq:infinite}
{\mathbf D}=
\left(\begin{array}{cccc}
{\bf D}_0 & 0 & 0 & \cdots \\
{\bf D}_1 &  {\bf D}_0 &0 &  \\
{\bf D}_2 & {\bf D}_1 & {\bf D}_0 &  \\
\vdots &  &  & \ddots
\end{array}\right),
\end{equation}
which is a formulation of the dynamics we will need in the following.

Finally, note that if we focus on promoter dynamics alone, the state of the system is specified by the vector ${\bf p}_t =\sum_{M}{\bf p}_t(M)$.
The corresponding master equation has the generator ${\mathcal D} = \sum_{n} {\bf D}_n$. In this case, the dynamics will be referred to as promoter-only dynamics, whereas the dynamics in \eref{eq:master} is the full system dynamics. In the following, unless otherwise stated, we will assume that the matrix ${\mathcal D}$ is irreducible.

\emph{Large deviations.---} 
Large deviation theory for Markov processes specifies the relative likelihood to observe large fluctuations in trajectory observables~\cite{Touchette2013,Chetrite2013}.
We will be interested in the fluctuations of the class of trajectory random variables 
\begin{equation}\label{eq:rvA}
A_t= {\sum_{j \to i} g_{ij}^{0}}   +  {\sum_{(i,M) \to (i,M+n)} g_{ii}^{n}}   +    {\int_{0}^{t} ds\,  f_{i(s)}},
\end{equation}
where the sums extend over all transitions along a random realization of the BMAP, with each transition weighted by the parameters $g$ and the time spent in each state weighted by $f$.
Notable examples are the promoter activity, that is the number of mRNAs/proteins produced up to time $t$ ($f_i=0$ and $g_{ij}^{n} = n ~\delta_{ij}$), or the total time spent in promoter state $i$ ($f_j= \delta_{ij}$ and  $g_{ij}^{n}= 0$).

Now, the law of large numbers tells us that for large $t$, the \emph{rate} $a=A_{t}/t$, will approach its average value ${\bar a}=\lim_{t\to\infty}A_t/t$ with near unit probability.
However, rare large deviations from this mean value are possible, and can be quantified, because the probability density of $a$ satisfies a large deviation principle for large $t$~\cite{Touchette2009,Ellis1985},
\begin{equation}\label{LDP}
P(A_t=at)\sim e^{-tI(a)}
\end{equation}
with large deviation rate function $I(a)$.
One can view this as an extension of the central limit theorem, quantifying not just the Gaussian fluctuations about the typical value (given by the minimum of $I$), but also the relative likelihood of rare fluctuations.

The theory of large deviations establishes that one way of obtaining the rate function $I(a)$ is by 
evaluating the (scaled) cumulant generating function (SCGF)~\cite{Ellis1985,Touchette2009}
\begin{equation}\label{eq:psi}
\psi(\lambda) = - \lim_{t\to\infty}\frac{1}{t}\ln \langle e^{-\lambda A_{t}}\rangle_t,
\end{equation}
where $\langle\cdot\rangle_t$ denotes an average over full system trajectories of duration $t$.
The rate function can then be recovered using the Legendre-Fenchel transform~\cite{Ellis1985,Touchette2009}
\begin{equation}\label{Legendre}
I(a)=\sup_{\lambda}(\psi(\lambda)-\lambda a).
\end{equation}
The function $I(a)$ ($\psi(\lambda)$) is the nonequilibrium analog of the entropy (free energy) in 
equilibrium systems. 
It is important to note that \eref{Legendre} only gives the convex hull of $I$, that is the smallest convex set that encompasses $I$~\cite{Touchette2009}. 
When $I$ is not strictly convex, that is having multiple local minima or having a linear part,
the SCGF $\psi$ is not differentiable everywhere, and is characterized by a nonanalyticity such as a kink.  
The appearance of such a structure in $\psi$ is called a \emph{dynamical phase transition}~\cite{Touchette2009,Garrahan2007,Vaikuntanathan2014} in analogy with the nonanalytic behavior of the free energy characterizing phase transitions in equilibrium statistical mechanics. 

A key insight from large deviation theory is that $\psi(\lambda)$ can be obtained from the largest eigenvalue of a modified or twisted generator matrix (cf.~\eref{eq:infinite})
\begin{equation}
\tilde{\mathbf D}(\lambda)=
\left(\begin{array}{cccc}
\tilde{\bf D}_0(\lambda) & 0   &\cdots \\
\tilde{\bf D}_1(\lambda) &  \tilde{\bf D}_0(\lambda)   \\
\vdots &  &   \ddots
\end{array}\right),
\end{equation}
with elements~\cite{Jack2010,Chetrite2013}
\begin{equation}
\tilde{D}_0^{ij}(\lambda)= 
\left\{\begin{array}{cc}-\sum_j\alpha_{ij}-k_j - \lambda f_j, & i=j \\
\alpha_{ij} e^{-\lambda g_{ij}^{0}}, & {i\neq j}\end{array}\right.; \quad \tilde{D}_n^{ij}(\lambda)=k_j b_j(n) e^{-\lambda g_{jj}^{n}}\delta_{ij}.
\end{equation}
To obtain $\psi$ from the largest eigenvalue of ${\tilde {\mathbf D}}(\lambda)$, we have to diagonalize an infinite-dimensional matrix. As we will show, the structure of BMAPs greatly simplifies this calculation, and reduces the computation to finding the dominant eigenvalue of an $N$-dimensional matrix, where $N$ is the number of promoter states.
 
The basic idea is to investigate the generating function $G_t(\lambda)=\langle e^{-\lambda A_t}\rangle_t$ for $A_t$, from which, in the long-time limit, the SCGF can be obtained as $\psi(\lambda)= - \lim_{t\to\infty}(1/t)\ln 
 G_t(\lambda)$ (by definition \eref{eq:psi}).
 We can express $G_t(\lambda)$ in terms of the joint probability at time $t$ to be at $(i,M)$ having accumulated $A$ of the trajectory observable, ${\bf p}_t(M,A)$:
\begin{equation}\label{eq:genFuncB}
G_t(\lambda)=\langle e^{-\lambda A_t}\rangle_t =\int \rmd A\, \sum_M {\bf 1} \cdot {\bf p}_t(M,A)e^{-\lambda A} \equiv \sum_{M^\prime}{\bf 1}\cdot {\bf G}_t(M^\prime;\lambda),
\end{equation}
where we have introduced the state-dependent generating function ${\bf G}_t(M;\lambda)=\int dA\, {\bf p}_t(M,A)e^{-\lambda A}$.
The key insight from large deviation theory is that ${\bf G}_t(M;\lambda)$ is the solution of the twisted dynamics~\cite{Jack2010}
  \begin{equation}\label{eq:genFuncA}
 {\dot{\bf G}}_{t}(M;\lambda) = \sum_{M^\prime} {\bf \tilde{D}}_{M-M^\prime}(\lambda) {\bf G}_{t}(M^\prime;\lambda),
 \end{equation}
 and its long-time behavior is controlled by the largest eigenvalue of its generator, the infinite-dimensional matrix $\tilde{\bf D}$.

The structure of BMAPs allows us to simplify this calculation significantly.
To this end, let us solve for the long-time dynamics of ${\bf G}_t$ by introducing $\bGamma_t(z;\lambda)=\sum_M  z^{M}{\bf G}_{ t}(M;\lambda)$, which from \eref{eq:genFuncA} evolves according to the simplified $N$-dimensional linear equation
\begin{equation}\label{eq:genFunEq}
\dot{\bGamma}_{ t}(z;\lambda) = \tilde{\mathcal D}(z;\lambda)\bGamma_{ t}(z;\lambda),
\end{equation}
where $\tilde{\mathcal D}(z;\lambda)=\sum_{n} z^{n} \tilde{\bf D}_n(\lambda)$.
Let us define $\tilde{\mathcal D}(\lambda) = \tilde{\mathcal D}(1;\lambda) = \sum_{n} \tilde{\bf D}_n(\lambda)$.
Notably,  $\tilde{\mathcal D}(\lambda)$ is the $N\times N$ generator for promoter-only twisted dynamics. 
The largest eigenvalue of $\tilde{\mathcal D}(\lambda)$ controls the long-time evolution of $\bGamma_t(z=1,\lambda)$ and thus gives us the SCGF.
Indeed, denoting the largest eigenvalue of $\tilde{\mathcal D}(\lambda)$ by $-\psi(\lambda)$ (anticipating the conclusion), and substituting the solution 
of \eref{eq:genFunEq} into \eref{eq:genFuncB} gives us
\begin{equation}
G_t(\lambda)=\langle e^{-\lambda A_t}\rangle_t={\bf 1}\cdot{\bGamma}_t(z=1;\lambda)\sim e^{-t\psi(\lambda)}.
\end{equation}
While the generator for the full system
dynamics is infinite dimensional, the SCGF is obtained from the dominant
eigenvalue of the $N$-dimensional matrix $ \tilde{\cal D}(\lambda)$ for
a model with $N$ promoter states, a substantial simplification.

\emph{Conditioning on rare fluctuations.---}
Suppose we want to know what typical trajectories of the system look like, conditioned on observing a (possibly rare) rate $a$ for the random variable $A_t$. 
Recent research in nonequilibrium statistical mechanics has shown that such information is encoded in the left eigenvector ${\mathbf{\tilde L}}(\lambda)$ corresponding to the dominant eigenvalue $\psi(\lambda)$ of the twisted generator ${\tilde {\mathbf D}}(\lambda)$~\cite{Chetrite2013,Vaikuntanathan2014,jack2015effective}.
Again, the structure of BMAPs leads to a significant simplification in determining this left eigenvector.
Denoting ${\tilde {\bf l}}_{\lambda}$ as the left eigenvector corresponding to $\psi$ for $\tilde{\mathcal D}(\lambda)$, we have that the infinite-dimensional left eigenvector ${\mathbf{\tilde L}}(\lambda)=\{{\tilde {\bf l}}(\lambda),\dots,{\tilde {\bf l}}(\lambda)\}$ is formed from repeated blocks of ${\tilde {\bf l}}(\lambda)$, due to the repeated block structure of ${\mathbf D}$ \eref{eq:infinite}.
Using this in combination with recent results for the driven process, we find that the conditioning-free Markov process that reproduces the statistics of the original process conditioned on $a$ has a generator with the same structure as \eref{eq:infinite}, with the corresponding transition rate matrices given by \cite{Chetrite2013,jack2015effective}
\begin{equation}\label{eq:twist}
\Delta_n^{ij}(\lambda^*)=l_i(\lambda^*) {\tilde D}_n^{ij}(\lambda^*)l_j^{-1}(\lambda^*)+\psi(\lambda^*)\delta_{ij}\delta_{n0},
\end{equation}
with $\lambda^*=-I'(a)$.
This surprising fact crucially depends on the class of random variables $A_t$ considered in \eref{eq:rvA}, which count promoter transitions with a weight that is independent of the number of mRNAs. 
Furthermore, given that the random variable $A_t$ is derived solely from dynamics related to the production of mRNAs/proteins, the results for the driven process are independent of the decay dynamics for mRNAs/proteins. In particular, we note that the results for parameters defining the driven model will be the same for the case where the decay rate is  $\mu = 0$ (which does not have a well-defined steady-state) 
and for the case for finite $\mu$ (which does have a well-defined steady-state).

By analyzing the equations determining the left eigenvector ${\tilde {\bf l}}_{\lambda}$ we find, remarkably, that the driven process corresponds to another BMAP with modified rates
\begin{equation}
\tilde\alpha_{ij}= l_i\alpha_{ij}l_j^{-1},\qquad \tilde{k}_i = k_i \sum_n b_i(n) e^{-\lambda g_{ii}^{n}}. 
\end{equation}
and renormalized burst distribution
\begin{equation}
\tilde{b}_i(n) = \frac{b_i(n) e^{- \lambda g_{ii}^{n}}}{ \sum_n b_i(n) e^{- \lambda g_{ii}^{n}}}.
\end{equation}

The above results, in combination with the equations determining the SCGF $\psi(\lambda)$ and rate function $I(a)$, constitute an analytical framework for characterizing rare events in a general class of stochastic models of gene expression.

\emph{Applications.---} In the following, we will focus on applications of the preceding  theoretical framework to the case of activity fluctuations ({\it i.e.}\ $f_i = 0$ and $g_{ij}^{n} = n \delta_{ij}$).

Let us consider rare events corresponding to low activity $a$ from a given promoter with a typical activity ${\bar a}$. 
This correspond to a fold-reduction in the mean activity of $r=a/{\bar a}$.
Let $\tilde{\psi}_{\lambda^*}$ with $\lambda^*=- I^\prime(a)$ denote the SCGF for the driven process. Then, by construction $a=\tilde\psi^\prime_{\lambda^*}(0)$.
As a result, the fold-change $r$ can be expressed in terms of the driven process as,
\begin{equation}\label{eq:req}
r = \frac{\tilde{\psi}'_{\lambda^*}(0)}{\psi'(0)}.
\end{equation}
However, we can flip this equation around and consider it as an equation for $\lambda^*$ in terms of the fold change $r$.
Once $\lambda^*$ is obtained by solving this equation, the driven
process for a given fold-change $r$ is completely determined. 

The driven process represents the most likely way the rare event in
activity fluctuations occurs and thus is an ideal choice for
generating distributions that can be used in importance sampling.
Recent work, using a control theory perspective, has also shown that the
driven process is the optimal way to to achieve the fold-reduction in
activity while minimizing the `distance', in the sense of relative
entropy, from the original process~\cite{Chetrite2015,jack2015effective}. Thus explicit characterization of
the driven process is also critical in designing regulatory strategies
for reducing promoter activity (thereby reducing mean mRNA/protein
levels) that give rise to processes that are closest to the original
process.

To illustrate the theoretical framework outlined above, we consider a widely used model of gene expression at the protein level~\cite{friedman2006linking,shahrezaei2008analytical,elgart2011connecting}: geometrically distributed bursts (with mean $b$) of proteins arriving according to a Poisson process with rate $k_{\rm m}$. In this case, the burst generating function is $g_{b}(z) = \frac{z}{b - z(b-1)}$ and correspondingly, we obtain that the SCGF $\psi(\lambda) = k_{\rm m} - k_{\rm m} g_b(e^{-\lambda})$. 
For a rare event corresponding to a $r$-fold reduction in mean activity, using \eref{eq:req} we see that the corresponding value of $\lambda^*$ is obtained by solving
\begin{equation}
\frac{e^{-{\lambda^*}}}{(b - e^{-{\lambda^*}}(b-1))^2} = r
\end{equation}
Interestingly, we find that the renormalized burst distribution for the driven process continues to be a geometric distribution with the corresponding mean given by
\begin{equation}
\tilde{b} = \frac{b}{b - e^{-{\lambda^*}} (b-1)}
\end{equation}
This indicates that the optimal way to reduce activity
involves arrival of geometric bursts with a reduced mean and a
modified arrival rate. Interestingly, in previous work \cite{jia2010post}, we have
shown that post-transcriptional regulation (e.g by miRNAs) can alter
bursts of protein expression by reducing the mean but keeping the
burst distribution geometric.  These results indicate that
that optimal control (in the sense discussed above) can be achieved
using a combination of transcriptional regulation and
post-transcriptional regulation by miRNAs.

\begin{figure}[htb]
\centering
\includegraphics[scale=.4]{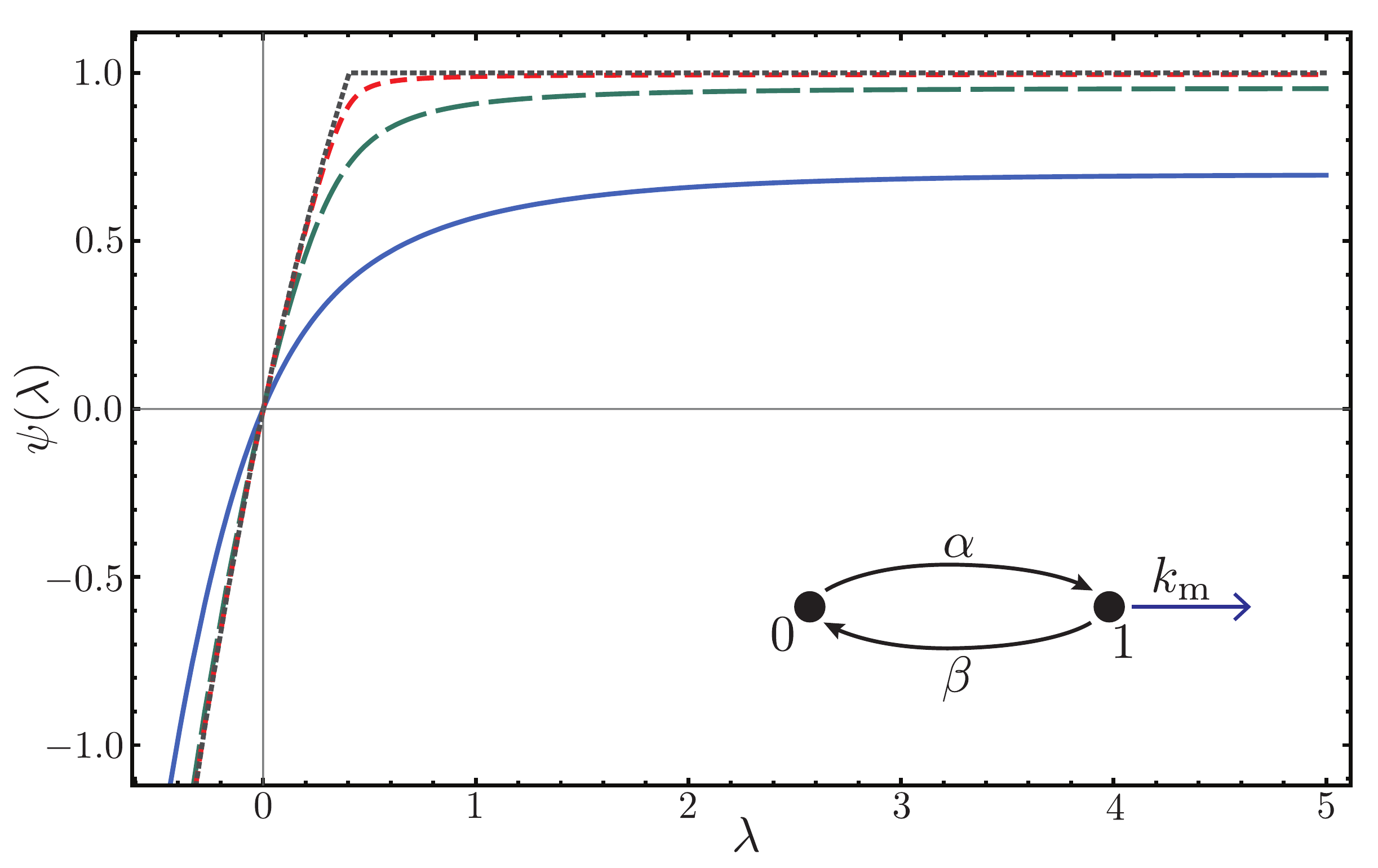}
\caption{Cumulant generating function $\psi(\lambda)$ for the two-state promoter model pictured inset. Different lines correspond to decreasing $\beta$ with the limiting value $\beta\to 0$ given by the dotted line.  All graphs have $\alpha<k_{\rm m}$ and thus are in the nonanalytic phase for $\beta\to 0$. Values: $\alpha=1$, $k_{\rm m}=3$, $\beta=1,0.1,0.01,0$. }\label{fig:kink}
\end{figure}

As a second illustration, we consider the two-state promoter model~\cite{sanchez2013genetic} depicted in \fref{fig:kink} with burst size one. 
The SCGF for mRNA activity is readily obtained by diagonalizing the twisted transition rate matrix:
\begin{equation}
\psi(\lambda) = \frac{1}{2}\bigg(\alpha+\beta+k_{\rm m}(1-e^{-\lambda}) -\sqrt{4\alpha\beta+\left(\beta-\alpha+k_{\rm m}(1-e^{-\lambda})\right)^2}\bigg).
\end{equation}
Remarkably, $\psi$ develops a nonanalyticity or kink in either the $\alpha\to0$ or $\beta\to 0$ limit.
Specifically for $\beta\to0$, we plot  in \fref{fig:kink} 
\begin{equation}
\psi(\lambda) = \frac{1}{2}\left(\alpha+k_{\rm m}(1-e^{-\lambda})-|\alpha-k_{\rm m}(1-e^{-\lambda})|\right),
\end{equation}
which is nonanalytic for $\alpha<k_{\rm m}$.  
The observed kink in the SCGF is a signature of a dynamical phase transition, akin to the multiphase behavior observed in photon counting statistics of a quantum two-level system~\cite{Garrahan2010}.  Here, the phase transition occurs because, as the activity level corresponding to the large deviation is changed, there is a qualitative change in the fluctuations that give rise to the rare event. In particular, below a critical activity level, these fluctuations  involve long sojourns in the 0 (OFF) state  (with no activity) that correspond to a nonvanishing finite fraction of the total time $t$.  We do note, however, that the nonanalytic behavior only arises when the matrix for promoter-only dynamics ${\mathcal D}$ becomes reducible. A detailed characterization of fluctuations in this limit will be presented in future work.

In conclusion, we have presented a framework for the quantitative analysis of the probability of large deviations and for characterizing system fluctuations conditional on rare events during gene expression. 
The framework developed provides explicit analytical formulae determining the driven process for a general class of stochastic models corresponding to BMAPs. Our results demonstrate that the driven process corresponding to a BMAP is another BMAP with renormalized parameters.  
This property may also be present in a related class of renewal processes, which, while distinct, share a similar structure in the dynamics~\cite{Harris2017}.
Since BMAPs are used to model a wide range of applications in science and engineering ({\it e.g.}\ computer and communications networks ~\cite{cordeiro2011batch}), the results derived are expected to have diverse applications. 
In the context of models of gene expression, the results derived can be used to: (1) determine the probability of rare events corresponding to a broad class of random variables for general promoter models with arbitrary mRNA/protein burst distributions; (2) directly simulate rare system trajectories corresponding to a large deviation in the random variable of interest and to predict the corresponding mRNA/protein distributions conditioned on the rare event; (3) optimally control gene expression to achieve, for example,  a specific fold-regulation while minimizing deviation from the unregulated process; and (4) determine regions in parameter space corresponding to nonanalytic behavior of the SCGF $\psi(\lambda)$.
The results obtained have thus opened new avenues for analyzing rare events in gene expression and have 
multiple applications ranging from importance sampling to optimal strategies for cellular regulation of gene expression. 

\ack 
The authors acknowledge funding support from the NSF through Awards DMS- 1413111  and PHY-1307067 as well as the Gordon and Betty Moore Foundation through Grant GBMF4343. RVK would also like to acknowledge funding support from the NIH through grants 3U54CA156734-05S3
and 2U54CA156734-06A.

\section*{References}
\bibliographystyle{iopart-num}
\bibliography{RareEventsBib.bib} 

\end{document}